\documentclass[final,nomarks]{dmtcs-episciences}
\usepackage[utf8]{inputenc}
\usepackage{subfigure}

\usepackage[round]{natbib}

\usepackage{amsmath}
\usepackage{amsfonts}
\usepackage{amsthm}
\usepackage{algorithm}
\usepackage{algpseudocode}
\usepackage{hhline}
\usepackage{varwidth}
\usepackage{graphicx}
\usepackage{float}
\usepackage{hyperref}

\newtheorem{theorem}{Theorem}

\newtheorem{fact}{Fact}

\newtheorem{lemma}{Lemma}
\newtheorem{claim}{Claim}

\author[L. Gargano, A. A. Rescigno]{Luisa Gargano\affiliationmark{1}\thanks{This work was partially supported by project SERICS (PE00000014) under the NRRP MUR program funded by the EU - NGEU.}
  \and Adele A. Rescigno\affiliationmark{1}\footnotemark[1]}
\title[Spanning Trees  Minimizing Branching Costs]{Spanning Trees  Minimizing Branching Costs}
\affiliation{
Department of Computer Science, University of Salerno, Fisciano (SA), Italy}
\begin{document}
% This is only used if you are compiling for a volume before vol 25
% \publicationdetails{VOL}{2015}{ISS}{NUM}{SUBM}
% This is the new form of collecting the data, starting with vol 25

\publicationdata{vol. 27:2}{2025}{17}{10.46298/dmtcs.13949}{2024-07-17; 2024-07-17; 2025-05-29}{2025-06-04}

\maketitle

%----------------------------------

\def\nd{{\tt{nd}}}
\def\cw{{\tt{cw}}}
\def\mw{{\tt{mw}}}
\def\vc{{\tt{vc}}}
\def\tw{{\tt{tw}}}

\def\SS{{{\sc Subset-Sum}}}
\def\CBV{{{\sc CBV}}}
\def\MWBV{{{\sc MWBV}}}
\def\MBV{{\sc MBV}}
\def\Br{{\tt Br}}

\def\V{{\cal{V}}}
\def\HB{{H_{B_H}}}
\def\HBp{{\hat{H}_{B_H}}}
\def\Hx{{H_x}}
\def\Hxp{{\hat{H}_x}}
\def\AB{{A_{B_H}}}
\def\A{{\alpha}}

\def\Ht{{{G}}}
\def\Gt{{{H}}}
\def\nt{{{n}}}
\def\mt{{{m}}}
\def\B{{B_\Gt}}
\def\GB{{\Gt_{B_\Gt}}}
\def\GBp{{\Gt'_{B_G}}}
\def\Gx{{\Gt_x}}
\def\Gxp{{\Gt'_x}}
\def\AB{{A_{B_\Gt}}}
\def\A{{\alpha}}
\def\P{{\cal{P}}}

\def\H{{\hat{\Ht}}}
\def\G{{\hat{\Gt}}}
\def\gG{{\hat{G}}}
\def\n{{\hat{n}}}
\def\m{{\hat{m}}}

\def\type{{module index }}
\def\types{{module indices }}

%-------------------------------------

\begin{abstract}
  The Minimum Branch Vertices Spanning Tree problem aims to find a spanning tree $T$ in a given graph $G$ with the fewest branch vertices, defined as vertices with a degree three or more in $T$. This problem, known to be NP-hard, has attracted significant attention due to its importance in network design and optimization. Extensive research has been conducted on the algorithmic and combinatorial aspects of this problem, with recent studies delving into its fixed-parameter tractability.

In this paper, we focus primarily on the parameter modular-width. We demonstrate that finding a spanning tree with the minimum number of branch vertices is Fixed-Parameter Tractable (FPT) when considered with respect to modular-width. Additionally, in cases where each vertex in the input graph has an associated cost for serving as a branch vertex, we prove that the problem of finding a spanning tree with the minimum branch cost (i.e., minimizing the sum of the costs of branch vertices) is FPT with respect to neighborhood diversity.
\end{abstract}

\section{Introduction}
Let $G=(V,E)$ be an undirected connected graph  where $V$ is the set of vertices and $E$ is the set of edges. Given a spanning tree $T$ of $G$, 
a  {\em branch vertex}  is a vertex having  degree   three or  more in $T$. 
We denote by  $b(G)$  the smallest number of branch vertices in any spanning tree of $G$.
We study the following  problem:
\begin{quote}
{\sc Minimum Branch Vertices} ({\sc MBV})\\
\textbf{Instance:} A connected graph  $G=(V,E)$.\\
\textbf{Goal:} Find  a spanning tree of  $G$   having $b(G)$ branch vertices.
\end{quote}
By noticing that the only   tree  without branch vertices is a  path, we know that  $b(G) = 0$ if and only if $G$ is Hamiltonian.

\medskip
The problem of determining a spanning tree with a limited number of branch vertices, though fundamentally a theoretical question, originated from addressing challenges in wavelength-division multiplexing (WDM) technology within optical networks. WDM is a highly effective method for utilizing the full bandwidth potential of optical fibers, thereby meeting the high bandwidth demands of the Internet \cite{HGCT02}.

Multicasting involves the simultaneous transmission of information from a single source to multiple destinations. In Wavelength-Division Multiplexing (WDM) systems, light trees are crucial for efficiently managing multicast communications by duplicating and routing specific wavelength channels. Light-splitting switches in these systems separate individual wavelengths within an optical signal. Due to the high cost of light-splitting switches, it is essential to minimize the number of nodes equipped with these devices. This need gives rise to the Minimum Branch Vertices (MBV) problem, which seeks to reduce the number of light-splitting switches in a light tree \cite{GHHSV}.  
In Cognitive Radio Networks, as well as in 5G technologies that operate across a broad spectrum of frequencies, managing the switching costs associated with transitions between different service providers is of utmost importance. This is essential not only for minimizing delays but also for optimizing energy consumption \cite{GBA,SR}.  
 
 The Minimum Branch Vertices (MBV) problem has been extensively studied from both algorithmic and graph-theoretic perspectives. Most prior research has focused on establishing upper bounds for the number of branch vertices in the resulting tree, though these bounds were often not tight. \cite{GHSV}  proved that 
 determining whether a graph $G$  has a spanning tree with at most $k$   branch vertices is NP-complete, for each $k \geq 1$, even in cubic graphs. In the same paper, they provided an algorithm that finds a spanning tree with one branch vertex if every set of three independent vertices in $G$
 has a degree sum of at least  $|V(G)|-1$.
%Results for  the MBV  problem have been given 
 \cite{Salamon}   developed an algorithm that finds a spanning tree with $O(\log |V (G)|)$ branch vertices for graphs where each vertex has a degree of  $\Omega(n)$; moreover,  an approximation factor better than $O(\log |V(G)|)$ would imply that $NP \subseteq DTIME(n^{O(\log \log n)})$. Sufficient conditions for a connected claw-free graph to have a spanning tree with 
$k$ branch vertices are presented in \cite{MOY}.

Integer linear formulations of the MBV problem and its variants are discussed in \cite{CCGG,CCR,CGI},  including different relaxations and numerical comparisons of these relaxations. Hybrid integer linear programs for MBV are considered and solved with branch-and-cut algorithms in \cite{SLC}.
Decomposition methods for solving the MBV problem are explored in \cite{LMS,MSU,RSS}, while other heuristics are presented in \cite{Marm,SSR+,SSR}.

A complementary formulation called the Maximum Path-Node Spanning Tree (MPN), which aims to maximize the number of vertices with a degree of at most two, is investigated in \cite{CS}.
The authors prove that MPN is APX-hard and provide an approximation algorithm with a ratio of 6/11. Related gathering processes are considered in 
 \cite{BGR,BGR1,BGR2,GR,GR1}.

\subsection{Graph Partitioning.} 
In order to address the Minimum Branch Vertices (MBV) problem, we also consider two related graph partitionings  that can be of  their own interest.

A {\em spider} is defined as a tree having at most one branch vertex, which is termed the {\em center} of the spider if it exists, otherwise any vertex can serve as the center.
A {\em path-spider cover of a graph $G$} consists of one spider and additional vertex-disjoint paths whose union includes every vertex of $G$. We denote by $spi(G)$ the least integer $p$ such that $G$ has a path-spider cover with  $p-1$ paths. 
We  define and study the following problem:

\begin{quote}
{\sc Path-Spider Cover} ({\sc PSC})\\
\textbf{Instance:} A graph $G=(V,E)$.\\
\textbf{Goal:} Find a path-spider cover of $G$ with $spi(G)-1$  paths.
\end{quote}
We will denote by by $\P_{spi(G)}$
a path-spider cover of $G$ with $spi(G)-1$  paths.

Moreover, we will make use of the   Partitioning into Paths problem defined below. 
A {\em partition of a graph $G$ into paths} is a set of (vertex-)disjoint
paths of $G$ whose union contains every vertex of $G$. We denote by $ham(G)$ the least integer $p$ such that $G$ has a partition into $p$ paths. Clearly, $spi(G) \leq ham(G)$.
\begin{quote}
{\sc Partitioning into Paths} ({\sc PP})
\\
\textbf{Instance:} A graph $G=(V,E)$.\\
\textbf{Goal:} Find a partition of $G$ into $ham(G)$  paths.
\end{quote}
We will denote by $\P_{ham(G)}$ a partition of $G$ into $ham(G)$  paths.\\
Notice that  the {\sc Partitioning into Paths} problem  was already considered in    \cite{GLO}; however as originally defined,  it only asks for the value $ham(G)$, while we ask for the actual path partitioning of $G$.
Recently, the {\sc PP} problem and some its variants have been considered in \cite{FFMPR}.

\smallskip

When referring to a path $P$ in $G$, we denote its end-points as $f(P)$ and $s(P)$, distinguishing them as {\em the first and the second end-point of $P$}, respectively. Additionally, if $P$ represents a spider in $G$, either $f(P)$ or $s(P)$ is used interchangeably to denote  {\em the center of $P$}. 

\subsection{Our Results} \label{algo-mbv-1}
 In Section \ref{algo-mbv}, we present an FPT algorithm for \MBV\ parameterized by   modular-width. To this aim, we also design  FPT algorithms for {\sc PSC} and {\sc PP} parameterized by   modular-width; they are presented in Section \ref{triple}.
 
 \medskip
Additionally,  in Section \ref{algo-cbv}  we  direct our focus to a scenario where the cost associated with selecting a vertex as a branch vertex varies across the graph. Consequently, we delve into the case where each vertex $v\in V$ is assigned a specific  non-negative integer weight  $w(v)$ and the cost of a spanning tree $T$ is  $w(T) = \sum w(u)$, where the sum is over all  the  branch vertices of $T$. The goal of the weighted version of the \MBV\ 
is to find a spanning tree $T$ of the input graph that  minimizes $w(T)$.
For the weighted \MBV\ problem we present an FPT algorithm parameterized by neighborhood diversity.
An FPT algorithm for the case of uniform cost was given in \cite{GR23-sir}.

\def \tw {{\tt tw}}

\subsection{Parameterized Complexity.} Parameterized complexity is a refinement to classical complexity theory in which one takes into account not
only the  input size, but also other aspects of the problem given by a parameter $p$.
%\cite{DF,Niedermeier}. 
A problem  with input size $n$ and  parameter $p$ is called {\em fixed parameter tractable (FPT)} if it can be solved in time $f(p) \cdot n^c$, where $f$ is a computable function only depending on $p$ and $c$ is a constant.

It was recently proven that  MBV   is FPT when parameterized by treewidth \cite{BW}.
The algorithm presented in \cite{BW} runs in time  $O(4^{2\tw} \tw^{4 \tw +1} (\tw +1)\;  n)$, where \tw\ is the treewidth of the input graph.
On the other hand, it was shown in \cite{Fomin} that the problem is $W[1]$-hard when parameterized by  clique-width. 
Specifically, in \cite{Fomin} it was proven that the (MBV special case) hamiltonian path problem is $W[1]$-hard when parameterized by  clique-width. 

In this paper, we are interested in assessing the complexity of MBV when parameterized by modular width and, in the weighted case,   by neighborhood diversity.

\begin{figure}[ht]  
\centering
\includegraphics[width=5.5truecm,height=4.0truecm,keepaspectratio]{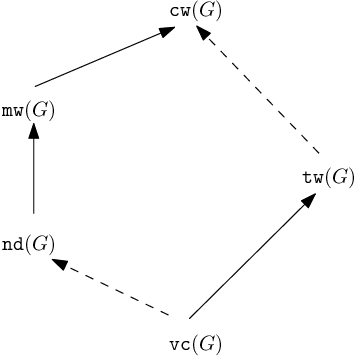}
\caption{A summary of the relations holding among some popular parameters. We use 
mw$(G)$, tw$(G)$, cw$(G)$, nd($G$), and 
vc$(G)$ to
denote modular-width, treewidth, cliquewidth, neighborhood diversity, and
minimum vertex cover of a graph $G$, respectively.
Solid arrows denote generalization, e.g., modular-width generalizes neighborhood diversity. Dashed arrows denote that the generalization may exponentially increase the
parameter.
}  \label{parameters}
\end{figure}

These 
graph parameters  could cover  dense
graphs but still allow FPT algorithms for the problems lost to clique-width.
See Figure  \ref{parameters} for a relation among the above  parameters.

\medskip
\noindent \textbf{Modular-width.}
Modular-width was introduced in \cite{GLO} and is defined as follows.

Consider graphs that can be obtained from an algebraic
expression that uses the following operations:
\begin{itemize}
\item[(O1)] \ {\em Create} an isolated vertex;
\item[(O2)] \ the {\em disjoint union} of 2 graphs denoted by $G_1 \oplus G_2$, i.e., $G_1 \oplus G_2$ is the
graph with vertex set $V(G_1) \cup V(G_2)$ and edge set $E(G_1) \cup E(G_2)$;
\item[(O3)] \  the {\em complete join} of 2 graphs denoted by $G_1 \otimes G_2$, i.e.,  $G_1 \otimes G_2$ is
the graph with vertex set $V(G_1) \cup V(G_2)$ and edge set $E(G_1) \cup E(G_2) \cup
\{ \{v,w\} : v \in V(G_1) \; \mbox{and} \; w \in V(G_2) \}$;
\item[(O4)] \  the {\em substitution} operation with respect to some graph $\Gt$ with vertex set $\{1,2, \ldots,n \}$
i.e., for graphs $G_1, \ldots,G_n$ the substitution of the vertices of $\Gt$ by
the graphs $G_1, \ldots,G_n$, denoted by $\Gt(G_1, \ldots,G_n)$, is the graph with vertex
set $\bigcup_{i=1}^n V(G_i)$ and edge set $\bigcup_{i=1}^n E(G_i)$ $\cup \{\{u, v\}  \ | \  u \in V(G_i),  v \in V(G_j), \{i,j\} \in E(\Gt)\}$. Hence, $\Gt(G_1, \ldots,G_n)$ is obtained from $\Gt$ by
substituting every vertex $i \in V(\Gt)$ with the graph $G_i$ and adding all edges
between the vertices of a graph $G_i$ and the vertices of a graph $G_j$ whenever
$\{i,j\} \in E(\Gt)$.
\end{itemize}
The {\em width} of an algebraic expression, that uses only the operations (O1)–(O4), is the maximum number of operands used by any occurrence of the operation (O4). The {\em modular-width} of a
graph $\Ht$, denoted $\mw(\Ht)$, is the least integer $m$ such that $\Ht$ can be obtained
from such an algebraic expression of width at most $m$. 
%\end{definition}

We recall that an algebraic
expression of width $\mw(\Gt)$ can be constructed in linear time \cite{TCHP}.

\begin{figure}[tb!]
	\centering
	\includegraphics[width=\textwidth]{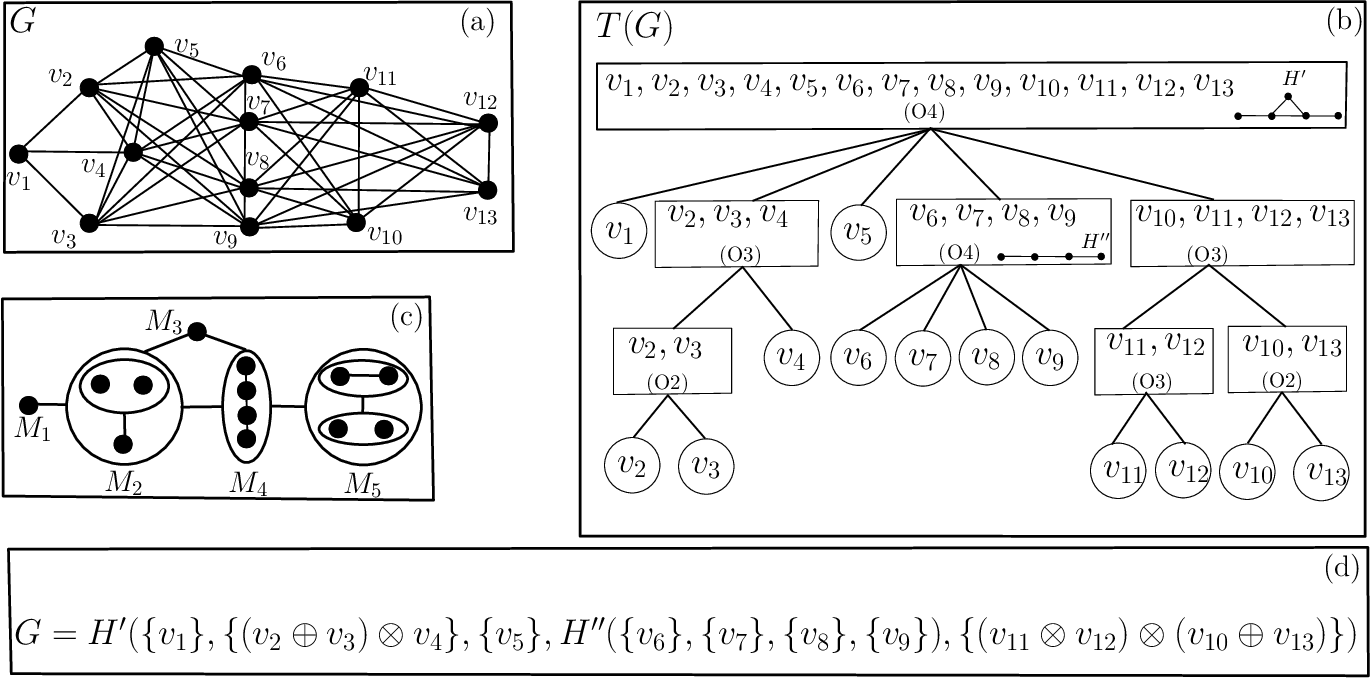}
		\caption{(a) A graph $G$. (b) The parse tree $T(G)$ associated with a decomposition of $G$ into modules. The width of the presented decomposition is $5$. (c) A representation of the decomposition of $G$ into modules. (d) An expression describing $G$ in terms of the operations (O1)-(O4).}
	\label{fig:imgMW}
\end{figure}

Given   a graph $G=(V,E)$, denote by $N(u)$ the set of neighbors of vertex $u\in V$. A {\em module} of a graph  $G$  is a set $M \subseteq V$ such that for all
$ u,v\in M, \  N(u)\setminus M=N(v)\setminus M.$
Operations (O1)-(O4) taken to construct a graph, form a parse-tree of the graph. A \textit{parse tree} of a graph $G$ is a tree $T(G)$  that captures the decomposition of $G$ into modules.   The  leaves  of $T(G)$  represent  the  vertices  of $G$ (operation (O1)).  The internal vertices of  $T(G)$ capture operations on modules: Disjoint union of its children (operation (O2)), complete join (operation (O3)) and substitution (operation (O4)). 
Figure \ref{fig:imgMW} depicts a graph $G$ and the corresponding parse tree.

\medskip
Given a graph $\Ht=\Gt(G_1, \ldots,G_n)$, we will  refer to the graphs $G_1, \ldots,G_n$ also as the modules of $\Ht$.
Notice that given the graph $\Ht=\Gt(G_1, \ldots,G_n)$,  by the operations O(1)-(O4), one has that all the vertices of  $G_i$ share the same neighborhood outside $G_i$; indeed,
\begin{equation}  \label{eq-edge}
\begin{array}{ll}{
\{\{u,v\}\, |\, u\in V(G_i), v\in V(G_j)\}\subseteq E(\Ht)}& \mbox{ if  $\{i,j\}\in E(\Gt)$}\\
{\{\{u,v\}\, |\, u\in V(G_i), v\in V(G_j)\}\cap E(\Ht)=\emptyset} & \mbox{ if  $\{i,j\}\notin E(\Gt)$}
\end{array}
\end{equation}
for each $i,j=1,\ldots,n$ with $i\neq j$.

\bigskip

\noindent \textbf{Neighborhood diversity.}
The neighborhood diversity parameter, was first introduced by Lampis in \cite{Lampis}.  It has then received much attention \cite{AAKK,ADNS,BR,CGRS,itp2,CorWAL,Gan,GR,GKK,TMKK}, also due to the fact that, contrary to other popular parameters,  it is computable in linear time.

Given   a graph $G=(V,E)$, two vertices $u,v\in V$  are said to have the same  
{\em type} if $N(v) \setminus \{u\} = N(u) \setminus \{v\}$.
The {\em neighborhood diversity} {\em \nd}$(G)$ of a graph $G$ is the minimum number $\nd$ of sets in a partition  $V_1,V_2, \ldots, V_\nd$, of the vertex set $V$, 
such that all the vertices in  $V_i$ have the same type, 
for  $i =1,\ldots,\nd$.

The family 
$\{V_1,V_2, \ldots, V_\nd\}$ is called  the {\em type partition} of $G$.
Notice that 
 each $V_i$ induces either a  clique or an  independent set in $G$.
Moreover, for each  $V_i$ and $V_j$ in the type partition,   either each vertex in 
$V_i$ is a neighbor of each vertex in $V_j$ or no vertex in $V_i$ 
has a neighbor  in $V_j$.
Hence, between each pair $V_i$ and $V_j$  there is either a complete bipartite graph  
or  no edge at all.
Starting from a graph $G$ and its type partition $\V=\{V_1, \ldots, V_\nd\}$,  
we can see each  element  of $\V$ as a   vertex of a new graph $H$,  called  the {\em type graph} of $G$,  with  \\
-   $V(H)=\{1,2,\cdots,\nd\}$ \\
-  $E(H)=\{\{x,y\} \ | \ x \neq y \mbox{ and for each $u\in V_x$, $v\in V_y$  it holds that $\{u,v\}\in E(G)$ }\}$.

Hence, a graph $G$ of neighborhood diversity $\nd$   can be seen as $G=H(G_1,\ldots, G_\nd)$, in which $G_i$ is the subgraph of $G$ induced by $V_i$ that is either a clique or an independent set, for each $i=1,\ldots,\nd$.

\section{The    MBV problem } \label{algo-mbv}

In this section we give an algorithm parameterized by \mw\  that  finds a spanning tree of graph $\Ht$ with $b(\Ht)$ branch vertices. Namely, we prove the following result.
\begin{theorem}
   {\sc Minimum Branch Vertices} parameterized by modular-width  is fixed-parameter tractable.  
\end{theorem}

Let $\Ht$ be the input graph.
Consider the parse-tree of an algebraic expression describing   $\Ht$, according to the
rules (O1)-(O4) in Section 1.3. We take a look at the  operation corresponding to the root: Operation (O1) is trivial and (O2) yields a disconnected graph, therefore we
suppose the last operation is either (O3) or (O4).
Hence, we can see the input graph as $\Ht=\Gt(G_1, \ldots,G_\nt)$ 
where  $\Gt$ is a connected graph with $2\leq \nt\leq \mw(\Ht)$ vertices and   $G_1, \ldots,G_\nt$ are graphs.

\medskip
The  algorithm that finds a spanning tree of an input graph $\Ht$  with $b(\Ht)$ branch vertices goes through the following steps 1) and 2).  
\begin{itemize}
\item[1)]  An FPT algorithm for PSC and PP parameterized by the modular-width of the input graph $\Ht$. Namely,
for each  vertex $\H=\G(\gG_1,\ldots, \gG_\n)$ of the  parse-tree of $\Ht$, we show how to compute the triple
$$\mbox{ $(ham(\H), spi(\H),|V(\H)|)$ together with $\P_{ham(\H)}$ and $\P_{spi(\H)}$,}$$
where $\P_{ham(\H)}$ is a partition of $\H$ into $ham(\H)$  paths  and 
$\P_{spi(\H)}$ is a path-spider cover of $\H$ with $spi(\H)-1$  paths.
\item[2)] Compute a spanning tree of  $\Ht=\Gt(G_1, \ldots,G_\nt)$ with $b(\Ht)$ branch vertices by using the  values, computed at step 1),  for the graphs   $G_1, \ldots,G_\nt$, that is, 
$$\mbox{ $(ham(G_i), spi(G_i),|V(G_i)|)$, \ $\P_{ham(G_i)}$, and $\P_{spi(G_i)}$,}$$
for   $i=1, \ldots, \nt$.
\end{itemize}
The  computation   in step 1)  is presented in Section  \ref{triple}. 
The computation in step 2), presented in this section,  is only done once, i.e., for the root vertex  of the parse tree, corresponding to the input graph $\Ht=\Gt(G_1, \ldots,G_\nt)$
whenever 
$\mbox{ $(ham(G_i), spi(G_i),|V(G_i)|)$, \ $\P_{ham(G_i)}$, and $\P_{spi(G_i)}$,}$
for   $i=1, \ldots, \nt$, are known.

\noindent
We start by giving  a characterization of  a  spanning tree with the minimum number of branch vertices in terms of the modular decomposition  of $\Ht$.
\begin{lemma} \label{bound}
Let  
$\Ht=\Gt(G_1, \ldots,G_\nt)$ be a connected graph.
There exists a spanning tree of $\Ht$ with $b(\Ht)$ branch vertices that has at most one branch vertex belonging to $G_i$ for each $i=1,\ldots, \nt$.
Hence, $b(\Ht)\leq \nt \leq \mw(\Ht)$.
\end{lemma}
\begin{proof}
Let $T$ be a spanning tree of $\Ht$ with $b(\Ht)$ branch vertices. 

Denote by $B$ the set of vertices of $\Ht$ that are branch vertices in $T$ and   by $N_T(v)$ the set of neighbors of $v$ in $T$, for any vertex $v$.
Assume that $|V(G_i)\cap B|\geq 2$ for some $1\leq i \leq n$. 
We show how to transform $T$ so to satisfy the lemma. 
The transformation consists of two phases.
\begin{itemize}
\item {\bf Phase 1.}
 For each $i=1,\ldots,n$, we denote by $B_i$  the set of branch vertices in $V(G_i)$ that have in $T$ at least two neighbors outside $V(G_i)$, that is, 
 $$B_i=\{ v \ | \ v\in V(G_i)\cap B \mbox{ and } |N_T(v)\cap (\cup_{j\neq i} V(G_j))| \geq 2\}.$$
In this phase we transform $T$ so that    $|B_i|\leq  1$, for each  $i$.
 We proceed as follows.
\begin{quote}
 For each  $i$ such that
$ |B_i|\geq 2,$
\begin{itemize}
\item[--] let $v$ be any node in $B_i$;
\item[--] for each $w\in B_i$ with $w\neq v$, consider the path connecting $v$ and $w$ in  $T$, say $v,\ldots, w', w$, and  modify $T$ as follows:
 For any $x\in (N_T(w)\setminus V(G_i))  \setminus \{w'\}$, substitute in $T$ the edge  $\{w,x\}$ by the edge $\{v,x\}$. 
\\
(Notice that this is possible by (\ref{eq-edge}) and implies  $B_i=\{v\}$). 
\end{itemize}
\end{quote}

\item {\bf Phase 2.}
We know that each $G_i$ contains at most one branch vertex with at least two neighbors outside $V(G_i)$, that is now $|B_i|\leq 1$ for each $i$. \\
Suppose  there exists $G_i$ such that $|V(G_i)\cap B|\geq 2$.
 We modify the spanning tree so that only one branch vertex remains among the vertices of $G_i$.
%to eliminate all those $v\in B$ such that \textcolor{red}{$|N_T(v)\setminus V(G_i)| \leq 1$}. 
We proceed as follows.
\begin{quote}
While there exists   $i$ such that $|V(G_i)\cap B|\geq 2$.
\begin{itemize}
\item[a)] Choose any   $j\neq i$ such that %$V(G_i)\times V(G_j)\subseteq E(\Ht)$ 
$\{i,j\} \in E(G)$
and let 
\begin{equation}
w\in 
\begin{cases} 
B_j & {\mbox{if $B_j=\{w\}$,}}\\ \nonumber
V(G_j) \cap B & {\mbox{if $B_j=\emptyset$ and $V(G_j) \cap B\neq \emptyset$}}\\
                {V(G_j) } & {\mbox{otherwise.}}
                \end{cases}\nonumber 
\end{equation}
\item[b)] For each  $v\in V(G_i)\cap B$  with $v \notin B_i$,
perform the following step.
\begin{itemize} 
\item[] Consider  the path connecting $v$ and $w$ in  $T$, say $v,v',\ldots w$, and   
 modify $T$ as follows: For any $x\in N_T(v)  \cap V(G_i)$ and $x \neq v'$, substitute in $T$ the edge  $\{v,x\}$ by the edge $\{w,x\}$. 
 \end{itemize}
 \end{itemize}
\end{quote}
Note that the aforementioned step is facilitated by (\ref{eq-edge}). It's important to emphasize that even if 
$w$ becomes a new branch vertex, we observe that  $|V(G_j)\cap B|= 1$. 
Furthermore,  $|V(G_i)\cap B|\leq 1$,  ensuring that the total count of branch vertices remains unchanged. Consequently, we have successfully derived a new spanning tree $T'$ with a set of branch vertices $B'$ such that $|B'|\leq |B|$. 
By repeating steps a) and b), one can construct the desired spanning tree of  $\Ht$ with at most one branch vertex in each $V(G_i)$.
\end{itemize}
\end{proof}

In the remaining part of this section, we present an algorithm that 
computes a spanning tree of $\Ht=\Gt(G_1, \ldots,G_\nt)$ with $b(\Ht)$ branch vertices, if $b(\Ht)> 0$. 
In Section \ref{ham} we deal with the case $b(\Ht)=0$, that is, we show how to find a Hamiltonian path of $\Ht$, if one exists.
The work in  \cite{GLO} focuses on verifying the existence of a Hamiltonian path in a graph; the proofs therein implicitly give a method for constructing such a path. 
We recall such a construction and show its relationship as a special case of our algorithm.

Using Lemma \ref{bound}, the algorithm proceeds by examining all subsets $\B\subseteq  \{1,\ldots, n\}$ with $|\B|\geq 1$, 
ordered by size. For each subset, it checks whether there exists a spanning tree of $\Ht$ with $|\B|$ branch vertices such that exactly one branch vertex belongs to each $V(G_i)$ with $i\in \B$ and none belongs to $V(G_i)$ for $i\not\in \B$.

 The identification of the spanning tree of $\Ht$ involves solving an Integer Linear Program (ILP) that utilizes the values  $ham(G_i), spi(G_i),|V(G_i)|$, for $i=1,\ldots,\nt$, and leverages property (\ref{eq-edge}).
Namely, if the ILP does not admit a solution for a subset $\B$, that subset is discarded. If a solution is found, we demonstrate how to use it along with the partitions of $G_i$ given by $\P_{ham(G_i)}$ and $\P_{spi(G_i)}$ to construct a spanning tree of $\Ht$ with exactly $|\B|$ branch vertices (noting that the subsets $\B$ are considered in increasing order of size).
The optimal spanning tree corresponds to the smallest subset $\B$ for which the ILP provides a solution.

\subsection{The Integer Linear Program} \label{ILP}

Let $\B \subseteq \{1,\ldots,\nt\}$, with  $|\B|\geq 1$. Construct a digraph 
$$\GB=(\{1,\ldots,\nt\}\cup \{s\},\AB),$$
where $s \not \in \{1,\ldots,\nt\}$ is an additional vertex that will be called the source.  $\GB$ is obtained from  $\Gt$ by replacing each edge  $\{i,j\}\in E(\Gt)$  by the two directed arcs $(i,j)$ and $(j,i)$, and then adding a directed arc $(s,r)$ where $r$ is  an arbitrary vertex in $\B$. Formally, 
$$\AB = \{(s,r) \} \cup \{(i,j), (j,i) \ | \ \mbox{there exists an edge between $i$ and $j$ in }E(\Gt)\}.$$
For sake of clearness, we will refer  to the vertices of $\Gt$ as {\em \types} 
and reserve the term vertex to those in $\Ht$.

We use the solution of the following  Integer Linear Programming (ILP) to select arcs of $\GB$ that will help to construct the desired  spanning tree in $\Ht$.

\begin{align}
&  x_{sr}  = 1 \\ 
&  \sum_{j : (j,i)\in \AB}  x_{ji}  \leq |V(G_i)|   \qquad\qquad  \qquad   \forall i \in \{1,\ldots,\nt\}  \\
&  \sum_{j : (j,i)\in \AB}  x_{ji}  \geq  spi(G_i)  \qquad\qquad \qquad   \forall i \in \B  \\
&  \sum_{j : (j,i)\in \AB}  x_{ji}  \geq  ham(G_i)  \qquad \qquad \qquad   \forall i \in \{1,\ldots,\nt\}\setminus\B  \\
& \sum_{\ell : (i,\ell)\in \AB}  x_{i\ell} -  \sum_{j : (j,i)\in \AB}  x_{ji} \leq 0  \qquad \   \forall i \in \{1,\ldots,\nt\}\setminus\B \\
&  y_{sr}  = \nt     \\
&  \sum_{j : (j,i)\in \AB}  y_{ji}  - \sum_{\ell : (i,\ell)\in \AB}  y_{i\ell}  = 1  \  \qquad   \forall i \in \{1,\ldots,\nt\}  \\
&  y_{ij} \leq \nt \ x_{ij}  \qquad\qquad \qquad \ \ \  \ \     \forall  (i,j) \in \AB\\
& y_{ij}, x_{ij} \in \mathbb{N} \qquad\qquad \qquad \ \ \ \ \ \  \forall  (i,j) \in \AB 
\end{align}

For each  arc $(i,j) \in \AB$, the non-negative decision variable $x_{ij}$ represents the load assigned to $(i,j)$. The load of the arc $(s,r)$ is set to 1. 
The total incoming load at \type $i\in \{1,\ldots,\nt\}$ must be at least  $spi(G_i)$ if
$i \in \B$ (ensuring the spider and all $spi(G_i)-1$ paths in $G_i$ are reached) and at least  $ham(G_i)$ $i \not \in  \B$ (ensuring all 
$ham(G_i)$ paths in $G_i$ are reached), and it must not exceed $|V(G_i)|$.   
Constraints (3), (4) and (5) enforce   these requirements.\\ 
Constraint (6) ensures that for any  $i \not \in \B$, the outgoing load is upper-bounded by the incoming load. \\
Constraints (7) and (8) implement a single commodity flow, with $s$  as the source and the other types as demand vertices.  For each  arc $(i,j) \in \AB$, the non-negative decision variable $y_{ij}$ represents the flow from  $i$ to  $j$.
\\
Each  $i \in \{1,\ldots,\nt\} $ has  demand of one unit, meaning the difference between inflow and outflow must be exactly one.
Meanwhile, the supply quantity at the source $s$ must be exactly $n$, to reach each of the \type in $\{1,\ldots, \nt\}$.
\\
Constraint (9) ensures that $y_{ij}=0$ whenever $x_{ij}=0$;  thus, if no load is assigned to $(i,j)$,  then   $j$ cannot be reached trough  $i$.

Given an integer solution $(y,x)$, if any,  to  the above ILP, the values of variables $y$ ensure that each \type $i \in \{1,\ldots,\nt\}$ is reachable from the source $s$. Consequently, by the construction of the digraph $\GB$, each \type $i \in \{1,\ldots,\nt\}$ is also reachable from \type $r$. Furthermore, due to the relationship between variables $x$ and $y$ (constraint (9)), we know that each \type $i \in \{1,\ldots,\nt\} $ receives incoming load from at least one of its neighbors.

\begin{claim} \label{path}
The subgraph $\Gx$ of $\GB$ with vertex set $\{1,\ldots,\nt\}$ and  arc set $\{(i,j)  | $ $ x_{ij} \geq 1\}$
 contains a directed path from $r$ to any other \type.
\end{claim}
We emphasize that the constraints involving variables  $y$ only guarantee the existence of a spanning tree in $\Gx$. A more sophisticated approach is needed to find a spanning tree of $\Ht$ that has exactly one branch vertex in  each $V_i$ with $i \in \B$.

\subsection{The spanning tree construction} \label{tree-construction}
Our algorithm  constructs a spanning tree $T$ of $\Ht$ with $|\B|$ branch vertices, one in each  $V(G_i)$  with $i\in \B$. To this aim, it uses the values of variables $x$ and  the path-spider cover $\P_{spi(G_i)}$ for $i \in \B$ and  the partition into disjoint paths $\P_{ham(G_i)}$  for $i \not \in \B$.

Denote by $In(i)$ the set of the \types for which there exist arcs in $\Gx$ toward  $i$, that is, $In(i)=\{j \ | x_{ji}\geq 1 \}$, 
and by
\begin{equation} \label{alpha}
\A_i = \sum_{j : j\in In(i)}  x_{ji}
\end{equation}
the number of vertices of $V(G_i)$ whose parent in $T$ is a vertex outside $V(G_i)$.\\
Let  $\P^i = \{P_1^i, P_2^i, \ldots, P_{\A_i}^i\}$ be
\begin{itemize}
    \item[$\bullet$]  the path-spider cover of $G_i$ obtained from those in $\P_{spi(G_i)}$ by removing $\A_i-spi(G_i)$ arbitrary edges  in case $i \in \B$ (notice that by constraint (4), it holds $\A_i \geq spi(G_i)$), or 
    \item[$\bullet$] the partition of $G_i$ into disjoint paths obtained from those in $\P_{ham(G_i)}$ by removing $\A_i-ham(G_i)$ arbitrary edges in case $i \not \in \B$ (notice that by  (5), it holds $\A_i \geq ham(G_i)$).
\end{itemize}
Furthermore, denote by
$$f(\P^i) = \{f(P_1^i), f(P_2^i), \ldots, f(P_{\A_i}^i)\}$$  
the sets of the first end-points  in the partition $\P^i$ and by
$$s(\P^i)=\{s(P_1^i), s(P_2^i), \ldots, s(P_{\A_i}^i) \}$$
the sets of the second end-points  in   $\P^i$.
In case $i \in \B$, we assume that $P_1^i \in \P^i$ is the spider and $f(P_1^i)=s(P_1^i)$ is the center in $P_1^i$.
\\
We also denote by
\begin{equation} \label{beta}
\beta_i = \begin{cases} 
 \sum_{\ell : i\in In(\ell)}  x_{i\ell} &  \text{if $i \not\in \B$} \\
 1 & \text{if $i \in \B$}
\end{cases}
\end{equation}
the number of vertices of $V(G_i)$, %among the above $\A_i$, 
that will be the parent of some vertex  in $\bigcup_{\ell : i\in In(\ell)}V(G_\ell)$.

\smallskip

Our algorithm ensures that the $\A_i$ vertices in $f(\P^i)$ are the vertices in $G_i$ whose parent in $T$ is located outside $V(G_i)$, and that $\beta_i$ vertices from  $s(\P^i)$ are selected to be  parents of vertices outside $V(G_i)$.  
Notice that by Claim \ref{path} ($\A_i \geq 1$)  and  constraint (6), it follows that $\A_i \geq \beta_i$ for each  $i =1,\ldots,\nt$. 

\begin{figure}[ht]
    \centering
    \includegraphics[width=14truecm,height=6.7truecm,keepaspectratio]{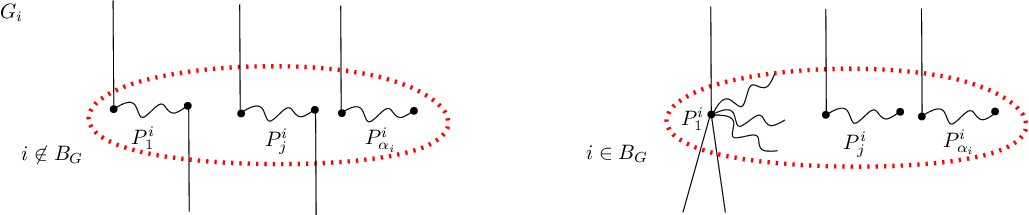}
    \caption{The vertices of graph $G_i$, grouped in the dotted circle, as partitioned in $\A_i$ disjoint paths if $i \not \in \B$ and in a spider plus $\A_i-1$ disjoint paths if $i  \in \B$. Vertex $f(P_j^i)$ is the only vertex in $P_j^i$ whose parent in $T$ is outside $G_i$ and vertex  $s(P_j^i)$ is the only vertex in $P_j^i$ that can have a children in $T$  outside $G_i$.}
    \label{fig:Gi}
\end{figure}

Figure \ref{fig:Gi} shows the partition of graph $G_i$ (whose vertices are grouped in the dotted circle) into $\A_i$ disjoint paths if $i \not \in \B$ and into a spider plus $\A_i-1$ disjoint paths if $i  \in \B$. 

The TREE algorithm iteratively constructs a spanning tree of $\Ht=\Gt(G_1, \ldots,G_\nt)$ by exploring unexplored vertices of $\Ht$, until all vertices are explored. It maintains a main subtree $T$ and a forest, with the roots of these trees progressively connected to  $T$ to form the spanning tree. The process halts when all vertices of $\Ht$ have been explored.

The algorithm maintains a set $R$, which contains the roots of trees with explored vertices that are waiting to be connected to the main tree $T$. The structure of the forest is described using the parent function
$\pi$.
At the beginning the set $R$ is empty. 
The exploration begins with the center $f(P_1^r)$ of the spider in the path-spider cover of $G_r$ 
(recall that $r \in \B$ by  construction ). The procedure EXPLORE$(f(P_1^r))$  constructs  the main tree $T$ rooted at $f(P_1^r)$ and marks all reached vertices as explored, adding them to the set $Ex$.
Clearly, for each explored vertex $v$, there is a path in $T$ that joins $f(P_1^r)$ to $v$.
The algorithm EXPLORE uses a queue $Q$ to  enqueue the explored vertices. 

However,  it is possible that some vertices remain unexplored (i.e., $V(\Ht)\setminus Ex \neq \emptyset$). In such scenarios, there exists a module $G_j$ where an explored vertex $w\in f(\P^j) \cap Ex$ exists alongside an unexplored vertex $u \in (f(\P^j) \setminus Ex)\setminus R$, capable of exploring at least one unexplored neighbor outside  $G_j$, that is,  $\beta_j \geq 1$
(the existence of such a set $V(G_j)$ is assured by Lemma \ref{cycle}). 
Utilizing  (\ref{eq-edge})   and  leveraging the knowledge that parents of vertices  in $f(\P^j)$ are outside $V(G_j)$,  the algorithm makes:
\begin{itemize}
\item  the parent of $w$ (recall that $w$ is explored) becomes the parent of $u$, and
\item 
 $w$,   the root of a subtree containing explored vertices, is added to  $R$ and removed from $Ex$. This action enables $w$ to be later explored and added, along with its subtree, to the main tree  $T$. 
  \item 
  EXPLORE$(u)$ is invoked to initiate a new exploration starting from $u$. 
 \end{itemize}

Notice that the algorithm updates  the forest by assigning the parent of $w$ to $u$.  Only later, after adding $u$ and some of its descendants,  the subtree rooted at $w$ is reintegrated into  the main tree $T$. This approach allows connecting new vertices in $V(\Ht)\setminus Ex$ to the main tree $T$.
The specific selection of $u$ and $w$ is designed to prevent scenarios where the algorithm could fail, ensuring that no arc can be added to  $T$ without either forming a cycle or introducing an additional branch vertex.
%(see Fig. \ref{tree-T} for an example).
 The process continues iteratively as long as there are unexplored vertices, i.e., $V(\Ht)\setminus Ex \neq \emptyset$.

The EXPLORE$(u)$ procedure  initiates exploration from  $u=f(P^j_k)$ along with the entire path or spider $P^j_k$
\footnote{Assume that when in the algorithm $P \in \P^j$ is explored (i.e.,  $Ex = Ex \cup P$),  the parent function $\pi$ is updated. For path, $\pi$ is set from $f(P)$ to $s(P)$, and for spiders, it is set from the center $f(P)=s(P)$ to the leaves.}.
Initially, only the vertex $s(P^j_k)$ is placed in $Q$. Subsequently, the main tree
$T$  is expanded.
If $u\neq f(P^r_1)$, the parent of $u$ is already a vertex in $T$, facilitating the construction of a subtree rooted at $u$
 that spans all newly explored vertices.

EXPLORE$(u)$ 
utilizes  the values of $\alpha_i$ and $\beta_i$ for each \type $i$,  defined initially   as in (\ref{alpha}) and (\ref{beta}), and the partition $\P^i=\{P_1^i, P_2^i, \ldots, P_{\A_i}^i\}$ of $G_i$.

The value 
$\A_i = \sum_{j : j\in In(i)}  x_{ji}$ counts the number of vertices of $V(G_i)$ that must have a parent  outside $V(G_i)$,  specifically $f(\P^i) = \{f(P_1^i), f(P_2^i), \ldots, f(P_{\A_i}^i)\}$. 
In particular, 
$x_{ji}$ vertices from $f(\P^i)$ need to be explored by vertices in $V(G_j)$, for $j \in In(i)$.
The value $\beta_i$ counts the number of vertices of $V(G_i)$ that need to explore other vertices in some other $V(G_\ell)$, for $\ell : i\in In(\ell)$.
In particular, 
\begin{itemize}
\item[-] if $i \in \B$ then exactly $\beta_i=1$ vertex in $V(G_i)$, that is $s(P^i_1)$ (i.e., the center of the spider $P^i_1$), becomes a branch vertex in $T$: it  is set as  the parent of $x_{i\ell}$ unexplored vertices in $f(\P^\ell)$ for each $\ell$ such that $x_{i\ell} \geq 1$ (i.e., $i\in In(\ell)$), and 
\item[-] if $i \not\in \B$ then  $\beta_i=\sum_{\ell : i\in In(\ell)}  x_{i\ell}$ vertices in $s(\P^i)=\{s(P_1^i), s(P_2^i), \ldots, s(P_{\A_i}^i) \}$ are chosen and each one becomes the parent of  one unexplored vertex in $f(\P^\ell)$.\\
Recall that, by the ILP constraints, we know that $\alpha_i \geq \beta_i$.
\end{itemize}

Only vertices in  $s(\P^i)$ for $1 \leq i  \leq \nt$ are enqueued in $Q$.
When a vertex $v \in s(\P^i)$ is dequeued from $Q$ in EXPLORE$(u)$    then the value of $\beta_i$ is decreased by one if $v$ explores (i.e., if $\beta_i \geq 1$).
In such cases, for each explored vertex $f(P^\ell_h)$ where $i \in In(\ell)$, the entire path 
 $P^\ell_h$ is also explored.

Additionally, the value $\alpha_\ell$ is  reduced  by the number of vertices in $s(\P^\ell)$ that $v$ explores, for $i\in In(\ell)$.  Therefore, at the start of each iteration of the while loop in  EXPLORE$(u)$  the value $\alpha_i$  represents the  number of vertices in $f(\P^i)$ that remain to   be explored while  $\beta_i$ indicates  the  number of vertices in $s(\P^i)$ that already need to explore.
It's noteworthy that  when a vertex $v \in s(\P^i)$ is dequeued from $Q$ in EXPLORE$(u)$, with $u \neq f(P^r_1)$, and  $\beta_i \geq 1$, the algorithm verifies if the neighbour $v'\in f(\P^\ell)$, which $v$ explores, is in $R$ (i.e., $v'$ is a root of a tree in the forest). If so,  $v'$ and its subtree are connected to the main tree $T$, as it has already been explored in  previous iterations.

\begin{algorithm}[!ht]
\caption{TREE$(\Ht, G_1, \ldots, G_\nt, r, \B)$}\label{alg_tree}
\begin{algorithmic}[1]
\State  $R=\emptyset$,  \ \ $B = \emptyset$,  \ \ $Ex = \emptyset$
\State $\pi(u)=nhil$ \ \ for each $u \in V(\Ht)$
%\State  Let $r \in \B$ 
\State EXPLORE$(f(P_1^r))$
\While{$V(\Ht)\setminus Ex \neq \emptyset$}
   \State - Let $G_j$ be any graph s.t. (($f(\P^j) \setminus Ex)\setminus R \neq \emptyset\neq f(\P^j) \cap Ex$) and $\beta_j\geq 1$ 
   \State - Let  $w \in f(\P^j) \cap Ex$ and $u \in (f(\P^j) \setminus Ex)\setminus R$
    \State  - Set $\pi(u)=\pi(w)$, \ \  $Ex=Ex - \{w\}$, \ \  $R=R \cup \{w\}$
      \State - EXPLORE$(u)$
\EndWhile
\State \textbf{return} $\pi$, \  $B$
\end{algorithmic}
\end{algorithm}

\begin{algorithm}[!ht]
\caption{EXPLORE$(u)$}\label{exp-u}
\begin{algorithmic}[1]
\State Let $Q$ be an empty queue
\State Let $u=f(P^j_k)$
\State  $Ex=Ex \cup P^j_k$ 
\State $Q.enqueue(s(P^j_k))$
\While{$Q \neq \emptyset$} 
     \State $v= Q.dequeue$   
     \State  Let $v \in s(\P^i)$
     \If{$i\notin \B$ and $\beta_i \geq 1$ }
              \State - Let $f(P^\ell_h) \in f(\P^{\ell})\setminus Ex$ for some $\ell$ s.t. $i\in In(\ell)$
             \State - $\pi(f(P^\ell_h)) = v$
             \If{$f(P^\ell_h) \not \in R$}  \State - $Ex=Ex \cup P^\ell_h$\State - $Q.enqueue(s(P^\ell_h))$
                 \Else \ \ $R=R\setminus \{f(P^\ell_h)\}$
             \EndIf
             \State -  $\A_\ell=\A_\ell-1$ \State - $\beta_i=\beta_i-1$
     \ElsIf{$i\in \B$ and $\beta_i=1$} 
        \State - $B=B \cup \{v\}$
        \For{each $\ell$ s.t. $i\in In(\ell)$}
        \State - Let $A_{i\ell}  \subseteq f(\P^\ell) -Ex$ s.t. $|A_{i\ell}|=x_{i\ell}$
        \State - $\A_\ell =\A_\ell - x_{i\ell}$,   \ \ 
         \For{each $f(P^\ell_h) \in  A_{i\ell}$}
          \State - $\pi(f(P^\ell_h)) = v$
          \If{$f(P^\ell_h) \not \in R$} \State -$Ex=Ex \cup P^\ell_h$ \State -$Q.enqueue(s(P^\ell_h))$
                 \Else \ \ $R=R\setminus \{f(P^\ell_h)\}$
             \EndIf
          \EndFor
        \EndFor
         \State  - $\beta_i=\beta_i-1$
    \EndIf
\EndWhile 
%\State \textbf{return} $Ex$
\end{algorithmic}
\end{algorithm}

\begin{lemma} \label{lemma-ur}
At the end of  EXPLORE$(f(P^r_1))$ the function $\pi$  describes a tree, rooted at $f(P^r_1)$, spanning  the set $Ex\subseteq V(H)$ of explored vertices. The vertices in $B\cap Ex$ are the  branch vertices.
\end{lemma}
\begin{proof} 
When EXPLORE$(f(P^r_1))$ is is invoked, the entire spider $P^r_1$  is  explored  (i.e. $Ex=Ex \cup P^r_1$ and it is added to $T$) and its center $s(P^r_1)$ is enqueued in $Q$. 
Subsequently, each time a vertex $v\in s(\P^i)$ is dequeued from $Q$ (recall, $v \in Ex$,  indicating it is an explored vertex), the algorithm can either terminate its exploration  (i.e., $\beta_i =0$) or explore unexplored neighbors of $v$ together with the path/spider it belongs to.
It can be shown that $v$  indeed has the necessary number of unexplored neighbors.
If $\beta_i =0$ then $v$ is a leaf in $T$, and only the case where $\beta_i \geq 1$ needs consideration.
If  $i \not \in \B$ then  $v$ has   $\beta_i \geq 1$  unexplored neighbors and one of them, say $f(P^\ell_h)$ for $i \in In(\ell)$, can be added to $T$ as child of $v$ .
If  $i \in \B$ then   $v$ is the first vertex of $V(G_i)$ to explore,  and 
 $x_{i\ell}$ vertices in $f(\P_\ell)$ are unexplored and can be  added to $T$ as children of $v$, for each $\ell$ such that  $x_{i\ell} \geq 1$.
Therefore,  $v$ becomes a branch vertex in $T$ and is added to $B$. 
Since $R = \emptyset$ (indicating no trees are in the forest),  each time a neighbor of $v$ is explored, such as $f(P^\ell_h)$, the entire path/spider $P^\ell_h$ is added to $T$ and $s(P^\ell_h)$ is enqueued in $Q$.
Then, every explored vertex has $f(P^r_1)$ as its ancestor, meaning the function $\pi$ defines a path form any explored vertex to $f(P^r_1)$. 
Since no  vertex can be  enqueued in $Q$ more than once (as each enqueued vertex is also marked as explored),  the function $\pi$ ensures that no cycles are created.
\end{proof}

Fix any iteration  of the while loop  in algorithm TREE and 
 define $\Gxp$ as the subgraph  of $\Gx$ (refer to Claim \ref{path}) containing the arc $(i,j)$ if, at the beginning of the while loop,  fewer than $x_{ij}$ vertices in $f(\P^j)$ have been assigned a parent in $s(\P^i)$.

\begin{lemma} \label{Hx}
Let $j$ be any \type in $\Gxp$. The following properties hold for $\Gxp$:
\begin{itemize}
\item[(a)]  If $f(\P^j) \subseteq Ex$ then the \type $j$  is  isolated  in $\Gxp$.
\item[(b)] The \type $r$  s.t. $f(P^r)$ contains the root $f(P^r_1)$ of $T$ has no outgoing arcs in $\Gxp$.
%\item[(c)]  $V_j \setminus Ex \neq \emptyset$  if and only if   $j$  has at least one  incoming  arc in $\Gxp$.
\item[(c)]  $f(\P^j)  \not \subseteq Ex $  if and only if   $j$  has at least one  incoming  arc in $\Gxp$.
\item[(d)] If  $f(\P^j) \cap Ex =\emptyset$   then  $j$  keeps in $\Gxp$   all the  incoming and outgoing arcs it has  in $\Gx$.
%\item[(e)] If $j\in \B$ and $V_j \cap Ex \neq \emptyset$ then $j$ has no outgoing arcs in $\Gxp$.
\item[(e)] If $f(\P^j) \setminus  Ex \subseteq R$ then $j$ has no outgoing arcs in $\Gxp$.
\end{itemize}
\end{lemma}
\begin{proof} 
%Property (a) follows by noticing that the relation $V_j \subseteq Ex$  is equivalent to say that  $\sum_t x_{t,j}=|V_j|=\sum_t x_{j,t}$.\\
Consider property (a). If each vertex in $f(\P^j)$ is explored, then 
it has a parent in $T$. Recalling that  $\sum_{t: t \in In(j)} x_{tj}=\A_j=|f(\P^j)|$, it  follows that $j$ has no incoming arc in $\Gxp$.
Moreover,
 procedure EXPLORE assures that  once a vertex in $f(\P^j)$ is explored (i.e. it has assigned a parent) then a vertex in $s(\P^j)$ will be assigned at least one  child as long as there are  vertices to be explored from  $V(G_j)$ (i.e., if $\beta_j \geq 1$).
In particular, if $j \not \in \B$ then since   $\sum_{t: j \in In(t)} x_{jt} =\beta_j=|s(\P^j)|$, it
 follows that   $x_{jt}$ vertices in $f(\P^t)$ have a parent in $s(\P^j)$, for each $t$ such that $j \in In(t)$. Hence  $j$ has no outgoing arc surviving in $\Gxp$.
 If, otherwise, $j \in \B$ then  vertex $s(P^j_1)$  has $x_{jt}>0$ children in $f(\P^t)$ for each $t$ such that $j \in In(t)$. Hence, also in this case  $j$  has no outgoing arc surviving in $\Gxp$.\\
Property (b) follows  from the observation that,  by construction, $f(P^r_1)$ is a branch vertex of $T$ and has  $x_{rt}$ children  in each $f(\P^t)$ such that $x_{rt}>0$, (i.e., $r \in In(t)$). \\
Property (c) follows  from the observation that $f(\P^j)  \not \subseteq Ex$  is equivalent to say that  
$$\sum_{t: t \in In(j)} x_{tj}=\A_j=|f(\P^j)|>|f(\P^j)\cap Ex|.$$
Property (d) follows from the observation that $f(\P^j) \cap Ex =\emptyset$ implies that $j$  still has an incoming  neighbour for each $t$ such that $x_{tj}>0$ and an outgoing neighbor for each $t$ such that $x_{jt}>0$.\\
%Property (e) follows   since, by procedure EXPLORE, the first  vertex $v \in V_j$  that is explored is a branch vertex; indeed, $v$ explores (e.g., gets as children)  $\sum_{t: t \in In(j)} x_{tj}$ unexplored vertices.\\
Property (e) follows  from the observation that when the algorithm TREE disconnects a vertex $w$ and adds it to $R$, vertex $w$ has already been assigned its child/children. Hence, if $f(\P^j)$ does not contain any unexplored vertex outside $R$   then $\beta_j$ has been decreased to 0. This means that all the $x_{tj}$ arcs from a vertex in $s(\P^t)$ to one in $f(\P^j)$ have been added to the forest, for each $j=1,\ldots, \nt$.
\end{proof}

\smallskip

We can prove the following results.
%The proof of the following Lemma is given in Appendix \ref{A}.
\begin{lemma}\label{cycle}
Let $Ex$ be the set of explored vertices at the beginning of any iteration of the while loop  in algorithm TREE. 
If $V(\Ht)\setminus Ex \neq \emptyset$ then there exists a \type $j$ such that $(f(\P^j) \setminus Ex)\setminus R\neq\emptyset\neq  f(\P^j) \cap Ex$ and $\beta_j\geq 1$.
\end{lemma}

\begin{proof}
By (a) and (c) of Lemma \ref{Hx}, we know that each \type $j$ in $\Gxp$ is either isolated or has at least one incoming arc. Hence, we focus on the subset of non-isolated \types. Since each of them has an incoming arc, $\Gxp$ contains a cycle.
Consequently,  each \type $j$ on such a cycle has an outgoing arc and satisfies $\beta_j\geq 1$.
Furthermore,   by (e) of Lemma \ref{Hx}, each \type $j$ on such a cycle satisfies
$(f(\P^j) \setminus Ex)\setminus R \neq \emptyset$.

We show now that at least one \type $j$ on the cycle has $f(\P^j)\cap Ex\neq \emptyset$.
Point  (b) of Lemma \ref{Hx} implies that  $\Gxp$ does not contain any path from $r$ to any \type on the cycle.
If we suppose that for each \type $j$ in the cycle $f(\P^j)\cap Ex=\emptyset$,  then (d)  of Lemma \ref{Hx} implies that  also $\Gx$ does not contain a   path from $r$ to $j$,  thus contradicting Claim \ref{path}.
\end{proof}

\begin{lemma}\label{lemma-u}
After each call of 
 EXPLORE$(u)$  the function $\pi$ describes a forest spanning  the vertices in $Ex \cup R$ of explored vertices and consisting of $|R|+1$ trees respectively rooted at $f(P^r_1)$ and at the vertices in $R$. The vertices in $B$ are the only branch vertices in the forest.
\end{lemma}
\begin{proof}
When  EXPLORE$(u)$ is called, the function $\pi$ describes a forest, spanning the current set  $Ex \cup R$ with roots  in $\{f(P^r_1)\}\cup R$, where $R \subset V(\Ht)\setminus Ex$.
According to  Lemma \ref{lemma-ur}, this is true the first time EXPLORE is called, that is,  after the call to  EXPLORE$(f(P^r_1))$ (at that time $R=\emptyset$).

We will prove that the claim holds at the end of each call to  EXPLORE$(u)$.
When  EXPLORE$(u)$ is invoked, $Q$ is initially empty. The vertex $u$ is  explored (i.e. added to $Ex$) and  enqueued in $Q$.
Then EXPLORE$(u)$ proceeds similarly to 
 EXPLORE$(f(P^r_1))$, dequeueing vertices from $Q$ and exploring their unexplored neighbors, thereby  constructing a subtree of the main tree $T$ rooted at $u$ described by the function $\pi$. 
The only difference from EXPLORE$(f(P^r_1))$ occurs when one of the explored vertices is $v'\in R$.
Vertex $v'\in R$ is removed from $R$ (see lines 14, 28), connected to the main tree $T$ through the function $\pi$,  and marked as explored like any other explored vertex. However $v'$ is not enqueued in $Q$ because it has already explored its neighbors; thus, $v'$ is connected to $T$ along with its subtree of explored vertices.
\end{proof}

\begin{lemma}
The algorithm TREE returns a spanning tree of $\Ht$, described by function $\pi$, with branch vertex set $B$.
\end{lemma}
\begin{proof}
By  Lemma \ref{lemma-ur} we know that algorithm TREE constructs a main tree $T$ through the procedure EXPLORE$(f(P^r_1))$, which is described by $\pi$. If $T$ does not span all the vertices in $V(\Ht)$ then, Lemma \ref{cycle} assures that 
the algorithm finds a graph $G_j$ with an explored vertex $w \in f(\P^j) \cap Ex$ and an unexplored vertex   $u \in (f(\P^j) \setminus Ex)\setminus R$. 
By disconnecting  $w$ (along its subtree) from the main tree $T$,  the algorithm allows $w$ to become one of the roots of trees in $R$. 
Additionally,  since the parent of $w$ in $T$ is a vertex outside $V(G_j)$ and, since  $u$ and $w$ share the same neighborhood outside $G_j$ (as indicated by (\ref{eq-edge})), the algorithm connects $u$ to the vertex that was the parent of $w$ in $T$ (thereby 
 connecting $u$ to $T$).
Since $u \not \in R$ and $\beta_j \geq 1$, the algorithm 
 starts a new exploration from $u$ (recall that $u \in f(\P^j) \setminus Ex$) by calling EXPLORE$(u)$.
 By Lemma \ref{lemma-u}, this allows  $T$ to be padded with the subtree rooted a $u$. The lemma follows by iterating the above procedure until no unexplored vertex exists in $V(\Ht)$.
\end{proof}
\subsection{The algorithm complexity} \label{complexity}
In summary, 
given
the triple $(ham(G_i), spi(G_i),|V(G_i)|)$ along with $\P_{ham(G_i)}$ and $\P_{spi(G_i)}$,  for each $i=1,\ldots,\nt$, the proposed method for constructing the spanning tree of $\Ht=\Gt(G_1,\ldots,G_\nt)$ works as follows.\\
For each  $\B \subseteq \{1, \ldots, \nt\}$ with $|\B|\geq 1$, selected in order of increasing size, 
 the  algorithm performs the following steps:
\begin{itemize}
\item  solve the corresponding ILP 
\item if a solution  exists for the current set $\B$,  use algorithm TREE to construct a spanning tree of $\Ht=\Gt(G_1,\ldots,G_\nt)$ with $|\B|$ branch vertices.
\end{itemize}

\cite{JR} have recently showed that the time needed to find a feasible solution 
of an ILP with $p$ integer variables and $q$ 
constraints is $O(\sqrt{q} \Delta)^{(1+o(1))q} + O(qp)$,
where $\Delta$ is the largest absolute value of any coefficient in the ILP. 
Denoted by $\mt$ the number of edges of $\Gt$,
our ILP has $q=3\nt+2\mt+1$ constraints, 
$p= 2(\mt+1)$ variables and $\Delta = \nt$.
Hence the time to solve it is $O(\nt\sqrt{\nt+\mt})^{(1+o(1))(3\nt+2\mt+1)} + O(\nt(\nt+\mt)).$
Using the solution $(y,x)$ of the ILP, the algorithm TREE returns the spanning tree of $\Ht$ in time $O(|V(\Ht)|^2)$.
Overall,  the algorithm requires  time
$$2^\nt [O(\nt\sqrt{\nt+\mt})^{(1+o(1))(3\nt+2\mt+1)} + O(\nt(\nt+\mt))]+O(|V(\Ht)|^2).$$
Recall that $n \leq \mw$, and therefore $m \leq \mw^2$.

\subsection{Optimality} \label{opt}
In this section we prove that the  spanning tree of $\Ht=\Gt(G_1,\ldots,G_\nt)$  with the minimum number of branch vertices is the tree that can be obtained by using the smallest set $\B\subseteq\{1,\ldots,\nt\}$ for which the ILP admits a solution.
Namely, we can  prove the following result.

\begin{lemma} \label{second}
Given the graphs $G_1,\ldots,G_\nt$ and $ham(G_i), spi(G_i), |V(G_i)|$ for each $i=1, \dots, \nt$.
Let $T$ be the spanning tree in $\Ht=\Gt(G_1,\ldots,G_\nt)$ with $b(G)$ branch vertices. Then there exists a set $\B\subseteq\{1,\ldots,\nt\}$ with $|\B|=b(G)$, for which ILP admits  a solution $(x,y)$.
\end{lemma}
\begin{proof}
Let $B$ be the set of branch vertices in $T$ and $|B|=b(G)$.
We show how to obtain from $T$ and $B$ an assignment of values to the variables in $x$ and $y$ that satisfy the constraint (2)-(9) of ILP.

By Lemma \ref{bound} we can assume that $|B \cap V(G_i)| \leq 1$ for each $i=1, \ldots, \nt$. 
Let $\B=\{i \ | \ B \cap V(G_i) \neq \emptyset, \ i=1 , \ldots, \nt\}$.
Choose any $r \in \B$ and let $u_r$ the branch vertex in $V(G_r)$. Root $T$ at $u_r$ and direct each edge in $T$ so that there is a path of directed arcs from $u_r$ to any vertex $u \in V(\Ht)\setminus \{u_r\}$. Let $A_T$ be the set of all the arcs in $T$.

We set $x_{sr}=1$ (satisfying constraint (2) of ILP), and for $i,j \in \{1 , \ldots, \nt\}$, 
$$x_{ij}=|\{(u,v) \ | \ (u,v)\in A_T, \ u \in V(G_i),\ v\in V(G_j)\}|.$$
Let $In(i)=\{j \ | \ x_{ji}\geq 1 \}$, for $i = 1 , \ldots, \nt$.
Since each vertex $u \in V(G_i)$ has a parent in $T$, we have the parent of any $u \in V(G_i)$ can be either a vertex in $V(G_i)$ or a vertex in $V(G_j)$ with $j \in In(i)$. This implies   that 
$$\sum_{j:(j,i) }  x_{ji}  \leq |V(G_i)|  \qquad\qquad\qquad \mbox{for each $i \in \{1, \ldots \nt\}$}.$$
satisfying constraint (3) of ILP.

Furthermore, if $V(G_i)$ does not contain a branch vertex then the tree $T$ induces at most $ham(G_i)$ disjoint paths in $G_i$. Hence, at least $ham(G_i)$ vertices in $V(G_i)$ have a parent in some $V(G_j)$ with $j \in In(i)$.
$$\sum_{j:(j,i) }  x_{ji}  \geq ham(G_i)  \qquad\qquad\qquad \mbox{for each $i \in \{1, \ldots,\nt\}\setminus \B$}$$
satisfying constraint (5) of ILP.\\
While, if $V(G_i)$  contains a branch vertex then $T$ induces a spider and at most 
$spi(G_i)-1$ disjoint paths in $G_i$. Hence, at least $spi(G_i)$ vertices in $V(G_i)$ have a parent in some $V(G_j)$ with $j \in In(i)$.
$$\sum_{j:(j,i) }  x_{ji}  \geq spi(G_i)  \qquad\qquad\qquad \mbox{for each $i \in \B$}$$
satisfying constraint (4) of ILP.

If $i \not \in \B$ then $V(G_i)$ does not contain branch vertices.
Hence, each vertex $u \in V(G_i)$ can be the parent of at most one vertex. Hence, 
$$\sum_{\ell : (i,\ell)}  x_{i\ell} \leq  \sum_{j : (j,i)}  x_{ji} \qquad\qquad\qquad \mbox{for each $i \in \{1, \ldots,\nt\}\setminus \B$}$$
satisfying constraint (6) of ILP.

To assign values to the variables $y$, we introduce the digraph $\Gt^x$ having
vertex set $\{1,\ldots,\nt\}$ and  arc set $\{(i,j) \ | \ x_{ij} \geq 1\}$.
Let $T_x$ be the tree rooted at $r$ obtained by a  BFS visit of $\Gt^x$.
For each $i \in  \{1 , \ldots, \nt\} \setminus  \{r\}$, let $p(i)$ the parent of $i$ in $T_x$. 
Pad $T_x$, adding arc $(s,r)$ (i.e., $p(r)=s$).
We set $y_{sr} =\nt$ (satisfying constraint (7) of ILP) and for $i \in  \{1 , \ldots, \nt\} \setminus  \{r\}$ we set
$$y_{ji}=\begin{cases}{\mbox{the number of vertices in the subtree of $T_x$ rooted at $i$}}&{\mbox{if $j =p(i)$}}\\
{0}&{\mbox{if $j \neq p(i)$}}
\end{cases}$$
Hence,
$$\sum_{j : (j,i)}  y_{ji} = y_{p(i)i} =  1+  \sum_{\ell : p(\ell)=i}  y_{i\ell} = 1+ \sum_{\ell : (i,\ell)}  y_{i\ell}$$
satisfying constraint (8) of ILP.

We notice that  the number of vertices in the subtree of $T_x$ rooted at $i$ is at most $\nt$, for each $i \in  \{1 , \ldots, \nt\}$. Moreover,  recalling that $x_{p(i)i} \geq 1$, we know that  $\Gt^x$ contains  $(p(i),i)$. Therefore, we get 
$$y_{p(i)i} \leq \nt \leq \nt \ x_{p(i)i}$$
satisfying constraint (9) of ILP since $y_{ji}=  0$  for each $j \neq p(i)$.
\end{proof}

\section{The triple and partition computation} \label{triple}
Following \cite{GLO},  we use a bottom-up dynamic programming approach along the parse-tree of an algebraic expression describing   $\Ht$,
to compute for every internal
vertex  $\H=\G(\gG_1,\ldots, \gG_\n)$ of the parse tree a record of data,  using those already computed for its children.
Specifically,  given
the triple $(ham(\gG_i), spi(\gG_i),|V(\gG_i)|)$,\, $\P_{ham(\gG_i)}$ and $\P_{spi(\gG_i)}$,  for each $i =1, \ldots, \n$,
we need to compute 
$$\mbox{the triple $(ham(\H), spi(\H),|V(\H)|)$ together with $\P_{ham(\H)}$ and $\P_{spi(\H)}$.}$$

Notice that in case $\gG_i$, for $i =1, \ldots, \n$, is a leaf in the parse tree (i.e., $\gG_i=(\{v\}, \emptyset)$ for some $v \in V(\H)$ $-$ operation (O1)) then $ham(\gG_i)= spi(\gG_i)=1$ and $\P_{ham(\H)} =\P_{spi(\H)}=\{v\}$.

Clearly, $|V(\H)|=\sum_{i=1}^\n |V(\gG_i)|$.
Below, we  show  how to compute $spi(\H)$ and   $\P_{spi(\H)}$, and also   $ham(\H)$ and   $\P_{ham(\H)}$.
%given in Appendix \ref{ham}.

For a graph $\H$ and an integer $\ell$, we denote by  $\H \otimes \ell$  the graph obtained from $\H$ by adding $\ell$ vertices and connecting them to every vertex in $\H$. Formally, $\H \otimes \ell$ has vertex set $V(\H) \cup \{v_1, \ldots, v_\ell\}$ and edge set $E(\H) \cup \{\{u,v_j \} \ | \ u \in V(\H), 1 \leq j \leq \ell  \}$.
Note that since $\H=\G(\gG_1, \ldots,\gG_\n)$, 
the graph $\H \otimes \ell$, for each $2 \leq \ell \leq |V(\H)|$,  is equal to the graph $\G'(\gG_1, \ldots,\gG_\n, I_{\ell})$ where 
$\G'$ is the graph obtained from $\G$ by adding
the vertex $\n+1$ (i.e., $V(\G')=\{1,\ldots,\n,\n+1 \}$) and making it adjacent to all the other vertices of $\G$ (i.e., $E(\G')=\{(i,\n+1) \ | \ 1 \leq i \leq \n\}$), and  $I_{\ell}$
is the independent set with $\ell$ vertices $\{v_1, \ldots, v_\ell\}$.

\subsection{Computing 
%$spi(\H)$
 \texorpdfstring{$spi(\H)$}{spi}
and 
%$\P_{spi(\H)}$
\texorpdfstring{$\P_{spi(\H)}$}{Pspi}}

In order to compute the values $spi(\H)$ and $\P_{spi(\H)}$,
 we first need a preliminary result.
\begin{fact} \label{fact1}
Let $\H$ be a graph and 
$$s(\H)= \min \{\ell \ | \ \H \otimes (\ell-1) \; \mbox{has a spanning spider with  center  in $\H$} \}.$$ 
Then $spi(\H) = s(\H)$.
\end{fact}
\begin{proof}
We first show that $s(\H) \leq spi(\H)$. Consider  $P_1,P_2,\ldots, P_{spi(\H)}$, the path-spider cover of $\H$, where $f(P_1)$ is the center of spider $P_1$ and, $f(P_i)$ denotes the first end-point of path $P_i$ for $i=2, \ldots, spi(\H)$.
Consequently, the graph 
$\H \otimes (spi(\H)-1)$ contains a spider with center $f(P_1)$, connecting  $f(P_1)$ to the vertex $i$, and then connecting vertex $i$ to $f(P_i)$ for each  for $i=1, \ldots, spi(\H)-1$.
Next, we establish that $spi(\H) \leq s(\H)$.
Let $S$ be a spider in $\H \otimes (s(\H)-1)$ with center $u \in V(\H)$. Removing  the vertices in $\{1, \ldots, s(\H)-1\}$ from $S$, we obtain a path-spider cover with center $u$ and $s(\H)-1$  pairwise disjoint paths. 
\end{proof}

\medskip

Recall that the graph $\H \otimes (\ell-1)$ is equal to  $\G'(\gG_1, \ldots,\gG_n, I_{\ell-1})$ and notice that  
$$  ham(I_{\ell-1})=spi(I_{\ell-1})=\ell-1 \mbox{ and } \P_{ham(I_{\ell-1})}= \P_{spi(I_{\ell-1})}= I_{\ell-1}.$$

 We can then take into account   the values $ham(\gG_i),\ spi(\gG_i), \ |V(\gG_i)|,$ and the sets $\P_{ham(\gG_i)}$ and $\P_{spi(\gG_i)}$, for each $i =1,\ldots,\n$. For each  $B_{\G'}=\{j\}$, for $j=1,\ldots, \n$,
we can follow the approach outlined in   Section \ref{ILP}  to determine the feasibility of the corresponding ILP. If feasible, according to the construction detailed in Section \ref{tree-construction},  it is possible to obtain a spider  of $\H \otimes (\ell-1)$ centered in $V(\gG_j)$. 
 
 The smallest integer $\ell$ for which the above holds determines $spi(\H)$ and provides a spider $T$ covering $\H \otimes (spi(\H)-1)$ with its center in $V(\H)$. The arguments outlined in Section \ref{complexity} allow us to derive  the time complexity of this computation (where, $\m$ represents the number of edges of $\G$)
 $$spi(\H) \; \n \;[O(\n\sqrt{\n+\m})^{(1+o(1))(3\n+2\m+1)} + O(\n(\n+\m))] + O(|V(\H)|^2).$$
It is evident that the subgraph of $T$ induced by $V(\H)$  returns the path-spider cover $\P_{spi(\H)}$ of $\H$.

\begin{theorem} \label{PSC}
{\sc Path-Spider Cover}  parameterized by modular-width is fixed-parameter tractable.    
\end{theorem}

\subsection{Computing 
 \texorpdfstring{$ham(\H)$}{ham}
and 
 \texorpdfstring{$\P_{ham(\H)}$}{Pham}} 
\label{ham}
Using an approach similar to the one in the proof of Fact \ref{fact1}, we can prove the following result.
\begin{fact} \label{fact2}
Let $\H$ be a graph and 
$$h(\H)= \min \{\ell \ | \ \H \otimes \ell \; \mbox{has a hamiltonian path with an end-point in $\{v_1, \ldots, v_\ell\}$}\}.$$ 
Then $ham(\H) = h(\H)$.
\end{fact}

\begin{proof}
We first show that $h(\H) \leq ham(\H)$. Consider $P_1,P_2,\ldots, P_{ham(\H)}$ the partition of $\G$ in disjoint paths, where $f(P_i)$ and $s(P_i)$ denote the first and second end-point of path $P_i$, respectively,  for $i=1, \ldots, ham(\H)$.
Consequently,  the graph 
$\H \otimes ham(\H)$ contains the hamiltonian path $P$ obtained connecting vertex $v_1$ to $f(P_1)$,   vertex $v_i$ to both $s(P_{i-1})$ and $f(P_{i})$ for each  for $i=2, \ldots, ham(\H)$. 

Next, we establish that $ham(\H) \leq h(\H)$.
Let $P$ be a hamiltonian path in $\H \otimes h(\H)$ with an end-point in $\{v_1, \ldots, v_{h(\H)}\}$. Removing  the vertices in $\{v_1, \ldots, v_{h(\H)}\}$ from $P$  we  obtain a partition of $\H$ in  $h(\H)$  pairwise disjoint paths. 
\end{proof}

\medskip

By Fact \ref{fact2}, the value  $ham(\H)$ is equal to
the smallest positive integer $\ell$ with $1 \leq \ell \leq|V (\H)|$ such that the graph 
\begin{equation}\label{H+l}
\mbox{$\H \otimes \ell$ has a hamiltonian path with an end-point  in  $\{v_1, \ldots, v_\ell\}$. }
\end{equation}

To determine if graph  $\H \otimes \ell$ has a hamiltonian path and  find it if it exists, we can follow the approach outlined in Section \ref{algo-mbv}.
Given that $\H \otimes \ell= \G'(\gG_1, \ldots,\gG_\n, I_{\ell})$, where $I_{\ell}$ is an independent set with $\ell$ vertices, we know  that  $ham(I_{\ell})=\ell$ and $\P_{ham(I_{\ell})}= I_{\ell}$. Therefore, with the values  $ham(\gG_i), |V(\gG_i)|$ and the set $\P_{ham(\gG_i)}$ available for each $i=1,\ldots,\n$,  we can construct the corresponding ILP as described in Section \ref{ILP} setting $B_{\G'}=\emptyset$ and $r=\n+1$ (i.e., $\gG_r=I_\ell$). If the ILP admits a solution, we can construct the hamiltonian path $P$ of $\H \otimes \ell = \G'(\gG_1, \ldots,\gG_\n, I_{\ell})$ using the method outlined in Section \ref{tree-construction}, choosing any vertex in $\gG_r=I_\ell$ as end-point (i.e., root). It is important to note that 
all proofs and constructions detailed in Section \ref{tree-construction} remains valid in this scenario  (i.e., $\G'(\gG_1, \ldots,\gG_\n, I_{\ell})$, $B_{\G'}=\emptyset$ and $\gG_r=I_\ell$).
Furthermore, as shown in Section \ref{opt}, it can be proved that if a hamiltonian path exists in  $\G'(\gG_1, \ldots,\gG_\n, I_{\ell})$ with an end-point in $I_\ell$, then exists a solution $(x,y)$ of the corresponding ILP (using the same arguments as in the proof of Lemma  \ref{second} with the hamiltonian path rooted at the end-point in $I_\ell$).

 The smallest integer $\ell$ for which (\ref{H+l})
occurs determines $ham(\H)$ and also provides the hamiltonian path $P$ of  $\H \otimes ham(\H)$ with one  end-point in  $I_{ham(\H)}$. 
The time complexity of this computation can be derived using the arguments presented in Section \ref{complexity}. We get
 $$ham(\H) \;[O(\n\sqrt{\n+\m})^{(1+o(1))(3\n+2\m+1)} + O(\n(\n+\m))] + O(|V(\H)|^2).$$
 
 Obviously, the subgraph of $P$ induced by $V(\H)$ will return the partition in $ham(\H)$ disjoint paths  of $\H$, $\P_{ham(\H)}$.

 \smallskip

We stress that  $ham(\H)=1$ iff the graph $\H$ has a hamiltonian path. 

\begin{theorem} \label{PP}
{\sc Partitioning into Paths}  parameterized by modular-width is fixed-parameter tractable.    
\end{theorem}

\section{The weighted \MBV\  problem} \label{algo-cbv}
Let $G=(V,E,w)$ be a weighted graph  where $w(u)$ is a positive integer representing the cost of choosing   $u$ as a branch vertex, for each $u \in V$. Given a spanning tree $T$ of $G$, let $B$ be the set of branch vertices in $T$. 
The {\em cost} of $T$ is the sum of the costs of the  vertices in $B$, that is $w(T) = \sum_{u \in B} w(u)$.
 Denote by $w(G)$  the minimum cost of any spanning tree of $G$,
 the weighted version of \MBV\ is formally defined as follows.
\begin{quote}
{\sc Minimum Cost Branch Vertices} ({\sc CBV})\\
\textbf{Instance:} A weighted undirected connected graph  $G=(V,E,w)$.\\
\textbf{Goal:} Find  a spanning tree $T$ of  $G$   having  cost $w(G)$.
\end{quote}. 
Trivially, in case $w(u)=1$ for each $u \in V$ then  \CBV\   becomes the problem of determining a spanning tree having 
the minimum number of branch vertices (\MBV).

We present an FPT algorithm for CBV parameterized by   neighborhood diversity.

Recall that a connected graph $G$ of neighborhood diversity $\nd$, with type partition $\V=\{ V_1, \cdots, V_\nd\}$ and type graph $H$ can be seen as $G=H(G_1,\ldots, G_\nd)$, where $G_i=G[V_i]$ is the subgraph of $G$ induced by $V_i$ that is either a clique or an independent set, for each $i=1,\ldots,\nd$.

\begin{lemma} \label{nd-bound}
Let $G=H(G_1,\ldots, G_\nd)$ be a connected graph with type partition  $\V=\{V_1,V_2,$ $ \ldots, V_\nd\}$ and type graph $H$. %, where $G_i=G[V_i]$ is the subgraph of $G$ induced by $V_i$.
For any spanning tree $T$ of $G$ with  cost $w(G)$ it holds:
\begin{itemize}
\item $|B \cap V_i |\leq 1$, 
\item if  $|B \cap V_i |= 1$ then $B \cap V_i$ consists of  a vertex having  minimum cost in $V_i$,
\end{itemize}
for each $i=1, \ldots \nd$. 
\end{lemma}
\begin{proof}
Let $T$ be a spanning tree of $G$ with branch cost $w(G)$.  
Denote  by $N_T(v)$ the set of neighbors of $v$ in $T$, for any vertex $v$, and by $d_T(v,w)$ the length of the path between $v$ and $w$ in $T$. 
Let $u_i$ be the vertex in $V_i$ having the minimum cost, for any $1 \leq i \leq \nd$.
We proceed by contradiction.

Assume that $|V_i\cap B|\geq 1$ and that no vertex $u_i$ of minimum cost $c(u_i)$ is in $V_i\cap B$, for some $1 \leq i \leq \nd$.
Let $v$ be any vertex in $V_i\cap B$. \\
In case $G_i$ is an independent set, let $v' \in N_T(v)$. We can modify $T$ to have a new spanning tree $T'$, as follows:   For any $x\in N_T(v)  \setminus \{v'\}$, substitute in $T$ the edge  $\{v,x\}$ by the edge $\{u_i,x\}$, then add the edge $\{u_i,v'\}$ in case it is not an edge in $T$.
Hence, the set of branch vertices $B'$ of $T'$ is $B'=(B\setminus \{v\}) \cup \{u_i\}$  and $w(T') =w(T)-w(v)+w(u_i) < w(T)$, having a contradiction.
\\
In case $G_i$ is a clique, we can modify $T$ to have a new spanning tree $T'$, as follows:   For any $x\in N_T(v)$, substitute in $T$ the edge  $\{v,x\}$ by the edge $\{u_i,x\}$, then add the edge $\{u_i,v\}$ in case it is not an edge in $T$. 
Again,  the set of branch vertices $B'$ of $T'$ is 
$B'=(B\setminus \{v\}) \cup \{u_i\}$  and $w(T') =w(T)-w(v)+w(u_i) < w(T)$, having a contradiction.

Assume now  that $u_i \in V_i\cap B$ and $|V_i\cap B|\geq 2$. Let $v\in V_i\cap B$ with $v\neq u_i$.\\
Consider first the case in which $G_i$ is an independent set.
Consider the path connecting $u_i$ and $v$ in  $T$, say $u_i,\ldots, v'', v', v$ .
We can modify $T$ to have a new spanning tree $T'$, as follows: For any $x\in N_T(v)  \setminus \{v'\}$, substitute in $T$ the edge  $\{v,x\}$ by the edge $\{u_i,x\}$, then, in case $d_T(u_i,v)\geq 3$, substitute in $T$ the edge  $\{v'',v'\}$ by the edge $\{u_i,v'\}$.
Hence,  the set of branch vertices $B'$ of $T'$ is $B'=B\setminus \{v\}$  and $w(T') < w(T)$, having a contradiction.
\\
Consider the case in which $G_i$ is a clique.
We can modify $T$ to have a new spanning tree $T'$, as follows: For any $x\in N_T(v)  \setminus \{u_i\}$, substitute in $T$ the edge  $\{v,x\}$ by the edge $\{u_i,x\}$, then,  in case $d_T(u_i,v)\geq 2$, add the edge $\{u_i,v\}$.
Again,  the set of branch vertices $B'$ of $T'$ is $B'=B\setminus \{v\}$  and $w(T') < w(T)$, having a contradiction.
\end{proof}

By using Lemma \ref{nd-bound}, we design  an algorithm that constructs a spanning tree of $G$ with branch cost $w(G)$, having  at most one branch vertex in each type set $V_i$ of the type partition $\V$ of $G$, and where, if $V_i$ contains a branch vertex then it is the vertex $u_i$ with minimum cost in $V_i$.

Our algorithm uses a simplified version of the algorithm presented in Section \ref{algo-mbv}.
We know that for each $i =1, \ldots,\nd$ the subgraph $G_i=G[V_i]$ induces either a clique or an independent set. As a consequence,  we have that:
\begin{itemize}
\item 
If $G_i$ induces a clique then\\
    -- $spi(G_i)=1$ and the path-spider cover of $G_i$, $\P_{spi(G_i)}$, is simply a star graph. We assume that the center of such a star is the vertex $u_i$  with minimum cost in $V_i$.\\
    -- $ham(G_i)=1$ and the partition of $G_i$, $\P_{ham(G_i)}$, consists of one path going through all the vertices in $V_i$.
\item 
If $G_i$ induces an independent set then \\
    -- $spi(G_i)=ham(G_i)=|V_i|$ and  $\P_{spi(G_i)}=\P_{ham(G_i)}=V_i$. 
    %consist of $|V_i|$ paths each formed by one vertex.
\end{itemize}
Hence, consider all the subsets of  $\{1,\cdots,\nd\}$ (with $|B_H|\geq 1$) arranged in ascending order of cost. For each non empty $B_H\subseteq \{1,\cdots,\nd\}$,
if we  choose  two vertices $r \in V(H)=\{1,\ldots, \nd\}$ and $s \notin \{1,\ldots, \nd\}$, and consider  directed graph $\AB = \{(s,r) \} \cup \{(i,j), (j,i) \ | \ \{i,j\} \in E(H) \}$,
the ILP in Section \ref{ILP} becomes:

\begin{align}
&  x_{sr}  = 1 \\ 
&  \sum_{j : (j,i)\in \AB}  x_{ji}  \leq |V_i|   \qquad\qquad  \qquad   \forall i \in \{1,\ldots,\nd\} \mbox{\ s.t. $G_i$ is a clique}  \\
&  \sum_{j : (j,i)\in \AB}  x_{ji}  = |V_i|   \qquad\qquad  \qquad   \forall i \in \{1,\ldots,\nd\} \mbox{\ s.t. $G_i$ is an ind. set}  \\
& \sum_{\ell : (i,\ell)\in \AB}  x_{i\ell} -  \sum_{j : (j,i)\in \AB}  x_{ji} \leq 0  \qquad \   \forall i \in \{1,\ldots,\nd\}\setminus B_H \\
%& \textcolor{blue}{\sum_{(j,i)}  x_{ji} = n-1}\\
&  y_{sr}  = \nd     \\
&  \sum_{j : (j,i)\in \AB}  y_{ji}  - \sum_{\ell : (i,\ell)\in \AB}  y_{i\ell}  = 1  \  \qquad   \forall i \in \{1,\ldots,\nd\}  \\
&  y_{ij} \leq \nd \ x_{ij}  \qquad\qquad \qquad \ \ \ \     \forall  (i,j) \in \AB\\
& y_{ij}, x_{ij} \in \mathbb{N} \qquad\qquad \qquad \ \ \ \ \ \  \forall  (i,j) \in \AB 
\end{align}

Whenever a feasible solution of the above ILP exists, the construction of the spanning tree $T$ of $G$ with branch vertices $B=\{u_i \ | \ i \in B_H \}$ follows the lines given in Section \ref{tree-construction} and algorithm TREE, once $\P_{spi(G_i)}$ and $\P_{ham(G_i)}$, for $i \in [\nd]$, are as described above.

The time complexity of the algorithm can be easily derived by considering that: The ILP is solved at most for each  $B_{H} \subseteq \{1,\cdots,\nd\}$ and, that it has at most $q=\nd^2 + 3 \nd +2$ constraints, at most $p=2(\nd^2+1)$ variables and $\Delta=\nd$; the algorithm TREE takes time $O(V(G)^2)$. By the results of Jansen and Rohwedderb \cite{JR}, overall the algorithm requires time
$$2^\nd(O(\nd^2)^{(1+o(1))(\nd^2+3\nd+2)} + O(\nd^4))+O(n^2).$$

It is easy to conclude that 
the  spanning tree of $G$  with the minimum  branch cost $w(G)$ is the tree that can be obtained by using the set $\B\subseteq\{1,\ldots,\nd\}$ having the least branch cost $w(B_H)$ for which the ILP admits a solution. Hence, we have the following result.

\begin{theorem}
   {\sc Minimum Cost Branch Vertices} 
   parameterized by neighborhood diversity  is fixed-parameter tractable.  
\end{theorem}

\section{Conclusions} 
We investigated the parameterized complexity of the Minimum Branch Vertices problem when parameterized by the modular width of the input graph, presenting an FPT algorithm. Additionally, we designed FPT algorithms for both the Path-Spider Cover and Partitioning into Paths problems, parameterized by  modular width. Furthermore, we presented an FPT algorithm parameterized by neighborhood diversity for the weighted version of the MBV problem,  the Minimum Cost Branch Vertices problem. The parameterized complexity of the Minimum Cost Branch Vertices problem remains an open question when parameterized by modular width.

\nocite{*}

\bibliographystyle{abbrvnat}

\bibliography{Branching-dmtcs-episciences} % Entries are in the refs.bib file

\end{document}